\documentclass[runningheads,fleqn]{svmult}

\usepackage{makeidx}   
\usepackage{graphicx}  
\usepackage{subeqnar}  
\usepackage{multicol}  
\usepackage{physproc}  
\makeindex             

\def\bG{{\bar{G}}}
\def\bV{{\bar{V}}}
\def\bep{\bar{\epsilon}_\K}
\def\de{\delta}
\def\De{\Delta}
\def\ep{\epsilon}
\def\hep{{\hat{\epsilon}}}
\def\k{{\bf{k}}}
\def\K{{\bf{K}}}
\def\kt{{\tilde{\k}}}
\def\mbG{{\hat{\bar{G}}}}
\def\mep{\hat{\epsilon}}
\def\mG{{\hat{G}}}
\def\mV{{\hat{V}}}
\def\mg{{\hat{g}}}
\def\mGa{{\hat{\Gamma}}}
\def\mS{{\hat{\Sigma}}}
\def\mt{{\hat{t}}}
\def\mT{{\hat{T}}}
\def\tk{\tilde{\bf{k}}}
\def\tx{\tilde{\bf{x}}}
\def\x{{\bf{x}}}
\def\X{{\bf{X}}}
\def\Gscr{{G_0}}

\begin{document}
\title*{Two Quantum Cluster Approximations}
\toctitle{Focusing of a Parallel Beam to Form a Point
\protect\newline in the Particle Deflection Plane}
%
%
\titlerunning{Two Quantum Cluster Approximations}
%
\author{Th.\ A.\ Maier\inst{1}
\and O.\ Gonzalez\inst{1,2}
\and M.\ Jarrell\inst{1}
\and Th. Schulthess\inst{2}}
\authorrunning{Th.\ Maier et al.}
%
%
\institute{University of Cincinnati\\
Cincinnati OH 45221, USA
\and Computational Material Sciences Group \\
Oak Ridge National Laboratory\\
Oak Ridge, TN  37831, USA}

\maketitle              

\begin{abstract}
We provide microscopic diagrammatic derivations of 
the Molecular Coherent Potential Approximation (MCA) and
Dynamical Cluster Approximation (DCA) and show that both 
are $\Phi$-derivable.  The MCA (DCA) maps the lattice onto a
self-consistently embedded cluster with open (periodic) 
boundary conditions, and therefore violates (preserves) the
translational symmetry of the original lattice.  As a consequence
of the boundary conditions, the MCA (DCA) converges slowly (quickly) 
with corrections ${\cal{O}}(1/L_c)$ (${\cal{O}}(1/L_c^2)$), where 
$L_c$ is the linear size of the cluster. However, local quantities,
when measured in the center of the MCA cluster, converge more quickly 
than the DCA result. These results are demonstrated numerically for 
the one-dimensional symmetric Falicov-Kimball model. 
\end{abstract}

\section*{Introduction}
	One of the most active areas in condensed matter physics
is the search for new methods to treat chemically disordered and correlated 
systems.  In these systems, especially in three dimensions or higher, 
approximations which neglect long ranged correlations are generally 
thought to provide a reasonable first approximation for many properties.

Perhaps the most successful of these methods are the Coherent Potential
Approximation (CPA) \cite{CPA} and the Dynamical Mean Field Approximation 
(DMFA) \cite{DMFA_Metzner,DMFA_MullerHartmann,DMFA_Pruschke,DMFA_Georges},
for disordered and correlated systems, respectively.  Although these
approximations have different origins, they are formally related.
Both are single site theories where non-local effects of
disorder and correlations are treated in a mean field
approximation. Diagrammatically,
both the DMFA \cite{DMFA_MullerHartmann} and the 
CPA \cite{DCA_Jarrell1} 
may be defined as theories which completely 
neglect momentum conservation at all internal vertices.  
When this principle is applied, the diagrammatic expansion for the 
irreducible quantities in each approximation collapses onto that of a 
self-consistently embedded impurity problem.

Many researchers have actively searched for a technique to restore 
non-local corrections to these approaches.  Here, we discuss just
two approaches which are fully causal and self-consistent: the 
Molecular Coherent Potential Approximation (MCA) \cite{MCPA,agonis} 
and the Dynamical Cluster Approximation 
(DCA) \cite{DCA_Hettler1,DCA_Hettler2,DCA_Maier1,DCA_Jarrell1}. 
Recently the Cellular Dynamical Mean Field 
Approach \cite{kotliar} was proposed for ordered correlated systems, 
while the Molecular Coherent Potential Approximation  has traditionally 
been applied to disordered systems.  Since both methods share a common 
microscopic definition we use the term MCA to refer to both techniques 
in the following.

While the MCA is traditionally defined in the real space of the 
lattice, the DCA is traditionally defined in its reciprocal space.  
In the MCA, the system lattice is split into a series of identical 
molecules.  Interactions between the molecules are treated in a 
mean-field approximation, while interactions within the molecule 
are explicitly accounted for.    In the DCA, the reciprocal space 
of the lattice is split into cells, and momentum conservation is 
neglected for momentum transfers within each cell while it is 
(partially) conserved for transfers between the cells.  These 
approximations share many features in common: they both map the 
lattice problem onto that of a self-consistently embedded cluster 
problem.  Both recover the single site approximation (CPA or DMFA) 
when the cluster size reduces to one and become exact as the cluster
size diverges.  Both are fully causal \cite{MCPA,DCA_Hettler2}, and 
provided that the clusters are chosen correctly \cite{DCA_Jarrell1}, 
they maintain the point group symmetry of the original lattice 
problem.  Here, we provide a microscopic diagrammatic derivation 
of both the MCA and the DCA, and explore their convergence
with increasing cluster size.

\section*{Formalism}
We will employ a diagrammatic formalism to derive the MCA and DCA,
assuming that a collection of electrons on a lattice, with Green 
function $G(\k)$ interact through an interaction $V(\k)$.  

\subsection*{From the Lattice to the Cluster}

Since our object is to define cluster methods, we divide the original lattice 
of $N$ sites into $N/N_c$ clusters (molecules), each composed of $N_c=L_c^D$ 
sites, where $D$ is the dimensionality.  $L_c$ need not be an integer, 
for example in a two-dimensional square lattice, a diamond cluster with 
$N_c=8$ will have $L_c=2 \sqrt{2}$.  However, care must be taken so that 
the clusters preserve the point group symmetry of the original lattice.
We use the coordinate $\tx$ to label 
the origin of the clusters and $\X$ to label the 
$N_c$ sites within a cluster, so that the site indices of the original 
lattice $\x=\X+\tx$.  The points $\tx$ form a lattice with a reciprocal 
space labeled by $\tk$.  The reciprocal space corresponding to the sites 
$\X$ within a cluster shall be labeled $\K$, with 
$K_\alpha=n_\alpha \cdot 2\pi/L_c$ and integer 
$n_\alpha$. Then $\k=\K+\tk$. Note that $e^{i\K\cdot \tx}=1$ since a 
component of $\tx$ must take the form $m_\alpha L_c$ with integer $m_\alpha$. 
\begin{figure}[htb]
\includegraphics[width=1.0\textwidth]{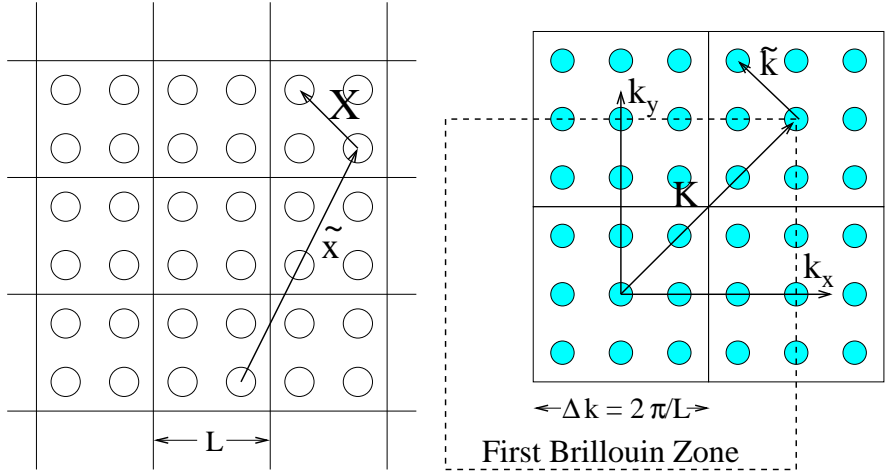}
\caption[]{Definitions of the parameters in real (left) and reciprocal (right) 
space.}
\label{fig:divide_x_k}
\end{figure}

The formal mapping between the full lattice problem and the cluster is 
accomplished by relaxing the condition of momentum conservation
at the internal vertices of the compact diagrams.  Momentum conservation 
at each vertex is described by the Laue function
\begin{equation}
\label{eq:Laue}
\De=\sum_\x e^{i\x\cdot(\k_1+\k_2+\cdots,-\k_1'-\k_2'-\cdots)}=
N\de_{\k_1+\k_2+\cdots,\k_1'+\k_2'+\cdots}\,,
\end{equation}
where $\k_1$, $\k_2$ ($\k_1'$, $\k_2'$) are the momenta entering (leaving) 
the vertex. M\"uller-Hartmann \cite{DMFA_MullerHartmann}
showed that the Dynamical Mean Field (DMF) theory may be derived by 
completely ignoring momentum conservation at each internal vertex by 
setting $\De=1$.  Then, one may freely sum over all of the internal momentum 
labels, and the graphs for the generating functional $\Phi$ and its 
irreducible derivatives,  contain only local propagators and interactions.

The DCA and MCA techniques may also be defined by their respective Laue 
functions.  In the MCA, we approximate the Laue function by
\begin{equation}
\label{eq:LMCA}
\De_{MC}=\sum_{\X}e^{i\X\cdot (\K_1+\tk_1+\K_2+\tk_2+\cdots
-\K_1'-\tk_1'-\K_2'-\tk_2'-\cdots)}\,.
\end{equation}
Thus, the MCA omits the phase factors $e^{i\tk\cdot\tx}$ resulting from the 
position of the cluster in the original lattice but retains the (far less 
important) phase factors $e^{i\tk\cdot\X}$ associated with the position 
within a cluster. In the DCA we also omit the phase factors  $e^{i\tk\cdot\X}$, 
so that
\begin{equation}
\label{eq:LDCA}
\Delta_{DC}=N_c\de_{\K_1+\K_2+\cdots,\K_1'+\K_2'+\cdots}\,.
\end{equation}
Both the MCA and DCA Laue functions recover the exact result when 
$N_c\to\infty$ and the DMFA result, $\Delta=1$, when $N_c=1$.

If we apply the MCA Laue function Eq.~\ref{eq:LMCA} to diagrams in 
$\Phi$, assuming that $V$ is a two-particle interaction then each 
Green function leg is replaced by the MCA coarse-grained 
Green function  (we have dropped the frequency
dependence for notational convenience)  
\begin{eqnarray}
\label{eq:cgMCA}
\bG(\X_1,\X_2;\tx=0)&=&\nonumber\\
&&\hspace*{-3cm}\frac{1}{N^2}\!\!\!\sum_{\stackrel{\K_1,\K_2}{\tk_1,\tk_2}}
e^{i(\K_1+\tk_1)\cdot\X_1}G(\K_1,\K_2;\tk_1,\tk_2)e^{-i(\K_2+\tk_2)\cdot\X_2}
=\nonumber\\
&&\hspace*{-3cm}\frac{N_c^2}{N^2}\sum_{\tk_1,\tk_2}G(\X_1,\X_2,\tk_1,\tk_2)\,,
\end{eqnarray}
or in matrix notation for the cluster sites $\X_1$ and $\X_2$
\begin{equation}
\label{eq:cgGMCA}
\hat{\bG}=\frac{N_c}{N}\sum_{\tk}\mG(\tk)\,,
\end{equation}   
since $\hat{\bG}$ can be chosen diagonal in $\tk_1, \tk_2$.
Similarly each interaction line is replaced by
\begin{eqnarray}
\label{eq:cgMCA2}
\bV(\X_1,\X_2;\tx=0)&=&\nonumber\\
&&\hspace*{-3cm}\frac{1}{N^2}\!\!\!\sum_{\stackrel{\K_1,\K_2}{\tk_1,\tk_2}}
e^{i(\K_1+\tk_1)\cdot\X_1}V(\K_1,\K_2;\tk_1,\tk_2)e^{-i(\K_2+\tk_2)\cdot\X_2}
=\nonumber\\
&&\hspace*{-3cm}\frac{N_c^2}{N^2}\sum_{\tk_1,\tk_2}V(\X_1,\X_2,\tk_1,\tk_2)\,,
\end{eqnarray}
or in matrix notation for the cluster sites $\X_1$ and $\X_2$
\begin{equation}
\label{eq:cgVMCA}
\hat{\bV}=\frac{N_c}{N}\sum_{\tk}\mV(\tk)\,.
\end{equation}   
The summations of the cluster sites $\X$ within each diagram remain 
to be performed.  $\hat{\bG}$ and $\hat{\bV}$ are propagators which
are truncated outside the cluster.  I.e., if the interaction $V$ is
non-local, $\hat{\bV}$ will include only interactions within, but not
between, clusters. Thus the inclusion of the phase factors 
$e^{i\tk\cdot\X}$ in the MCA Laue-function Eq.~\ref{eq:LMCA} leads 
directly to a cluster approach formulated in real space that violates 
translational invariance. Therefore the Green function and interaction
are functions of two cluster momenta $\K_1$, $\K_2$ or two sites 
$\X_1$, $\X_2$ respectively.

If we apply the DCA Laue function Eq.~\ref{eq:LDCA}, Green function 
legs in $\Phi$ are replaced by the DCA coarse grained Green function
\begin{equation}
\label{eq:cgGDCA}
\bG(\K)=\frac{N_c}{N}\sum_{\tk}G(\K,\tk)\,,
\end{equation} 
since Green functions can be freely summed over the $\tk$ vectors within 
a cell about the cluster momentum $\K$. 
Similarly, the interactions are replaced by 
the DCA coarse grained interaction
\begin{equation}
\label{eq:cgVDCA}
\bV(\K)=\frac{N_c}{N}\sum_{\tk}V(\K,\tk)\,.
\end{equation} 
As with the MCA, the effect of coarse-graining the interaction is to reduce
the effect of non-local interactions to within the cluster.  The resulting 
compact graphs are functionals of the coarse grained Green function 
$\bar{G}(\K)$ and interaction $\bV(\K)$, and thus depend on the cluster 
momenta $\K$ only.  For example, when $N_c=1$, only the local part of the 
interaction survives the coarse graining.  As with the MCA, within the DCA
it is important that {\em{both}} the interaction and the Green function are 
coarse-grained \cite{DCA_Hettler2}.  In calculations where a non-local 
interaction is not coarse-grained, poor results are obtained \cite{Biroli}.

\subsection*{From the Cluster to the Lattice}

To establish a connection between the cluster and the lattice we 
minimize the lattice free energy
\begin{equation}
F= -k_B T\left(
\Phi-\mbox{tr}\left[{\bf{\Sigma}} {\bf{G}}\right] 
+\mbox{tr}\ln\left[{\bf{G}}\right]\right)
\label{F_CA}
\end{equation}
where $\Phi$ is the generating functional composed of all closed compact
(single-particle irreducible) 
graphs, ${\bf{\Sigma}}$ is the lattice self-energy and ${\bf{G}}$ is the 
full lattice Green function.  The trace indicates summation over frequency, 
momentum and spin.  As discussed in many-body texts \cite{agd}, the additional 
free energy due to an interaction may be described by a sum over all closed 
connected graphs.  These graphs may be further separated into compact and 
non-compact graphs.  The compact graphs, which comprise the generating 
functional $\Phi$, consist of the sum over all skeletal graphs (those 
with no internal parts representing corrections to the single-particle
Green function).  The remaining graphs comprise the non-compact 
part of the free energy.  In the infinite-dimensional limit, $\Phi$ 
consists of only local graphs, with non-local corrections of order 
$1/D$.  However, for the non-compact parts of the free energy, non-local 
corrections over arbitrary lengths are of order one, so the local 
approximation applies only to $\Phi$.  

\begin{figure}[htb]
\includegraphics[width=0.5\textwidth]{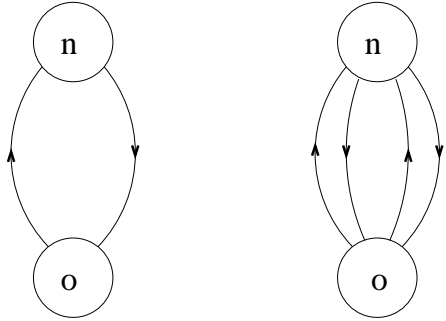}
\caption[]{Non-compact (left) and compact (right) non-local 
corrections to the free energy functional.  Here the upper (lower) 
circle is meant to represent a set of graphs which are closed except 
for the external lines shown, and restricted to site $n$ (the origin).}
\label{non_compact}
\end{figure}
To see this, consider the simplest non-local corrections to non-compact 
and compact parts of the free energy of a Hubbard-like model, illustrated 
in Fig.~\ref{non_compact}.  Here the upper (lower) circle is a set of 
graphs composed of intrasite propagators restricted to site $n$ (the 
origin).  Consider all such non-local corrections on the shell of sites 
which are $n$ mutually orthogonal unit translations from the origin.  
In the limit of high dimensions, there are  
$2^n D!/((D-n)! n!)\sim {\cal{O}} (D^n)$ such sites.  Since as 
$D\to\infty$, $G(r)\sim D^{-r/2}$ \cite{DMFA_Metzner}, the legs
on the compact correction contribute a factor ${\cal{O}}(D^{-2n})$
whereas those on the non-compact correction contribute ${\cal{O}}(D^{-n})$.
Therefore the compact non-local correction falls as $D^{-n}$ and is 
very short-ranged; whereas, the non-compact correction remains of order 
one, regardless of how far site $n$ is from the origin \cite{DCA_Karan1}.  
As we will see below, the essential approximation of the DCA and the 
MCA is to use the cluster propagators, which are accurate only for short 
distances, to construct various diagrammatic insertions. 
In high dimensions, or in finite dimensions when the Green functions fall 
exponentially with distance, this is a good approximation for 
the compact graphs which comprise $\Phi$, but a poor approximation
for the non-compact graphs \cite{DCA_Karan1}.

Thus, we will approximate the generating functional $\Phi$ with its cluster 
counterpart $\Phi_c$ by replacing 
the Laue function with either $\Delta_{DC}$ or $\Delta_{MC}$, but this
approximation will not be used in the parts of the free energy 
coming from non-compact graphs. The free energy then reads
\begin{equation}
F= -k_B T\left(
\Phi_{c}-\mbox{tr}\left[{\bf{\Sigma}} {\bf{G}}\right] 
+\mbox{tr}\ln\left[{\bf{G}}\right]\right)
\label{F_C}
\end{equation}
$F$ is stationary with respect to ${\bf{G}}$ when $\frac{\delta F}{\delta G}=0$.
This happens for the MCA if we estimate the lattice self energy as
\begin{eqnarray}
\label{eq:MDMFS}
\Sigma(\K_1,\K_2;\tk_1,\tk_2)&=&\nonumber\\
&&\hspace*{-3cm}\sum_{\X_1,\X_2} e^{-i(\K_1+\tk_1)\cdot\X_1}
\Sigma_{MC}(\X_1,\X_2) e^{i(\K_2+\tk_2)\cdot\X_2}\,.
\end{eqnarray}
Thus, the corresponding lattice single-particle propagator reads in matrix 
notation 
\begin{equation}
\label{eq:GXk}
\mG(\tk,z)=\left[zI-\mep(\tk)-\mS_{MC}(z)\right]^{-1}\,,
\end{equation}
where the dispersion $\mep(\tk)$ and self-energy $\mS_{MC}(z)$ are 
matrices in cluster real space with 
\begin{eqnarray}
[\mep(\tk)]_{\X_1\X_2}&=&\epsilon(\X_1-\X_2,\tk)\\\nonumber
&=&\frac{1}{N_c}\sum_\K e^{-\K\cdot(\X_1-\X_2)}\epsilon_{\K+\kt}
\end{eqnarray}
being the intracluster Fourier transform of the dispersion.
For the DCA, $\Sigma(\k)=\Sigma_{DC}(\K)$ is the proper 
approximation for the lattice self energy corresponding to $\Phi_{DC}$.
The corresponding lattice single-particle propagator is then given by 
\begin{equation}
G(\K,\tk;z) =\frac{1}{z-\ep_{\K+\tk}-\Sigma_{DC}(\K,z) } \,.
\label{G_DCA}
\end{equation}
Both the MCA and DCA are optimized when we equate the lattice and
cluster self energies.  A similar relation holds for two-particle
quantities.  Thus, with few exceptions \cite{exception}, only the 
irreducible quantities on the cluster and lattice correspond one-to-one.

\begin{figure}[htb]
\includegraphics[width=1.0\textwidth]{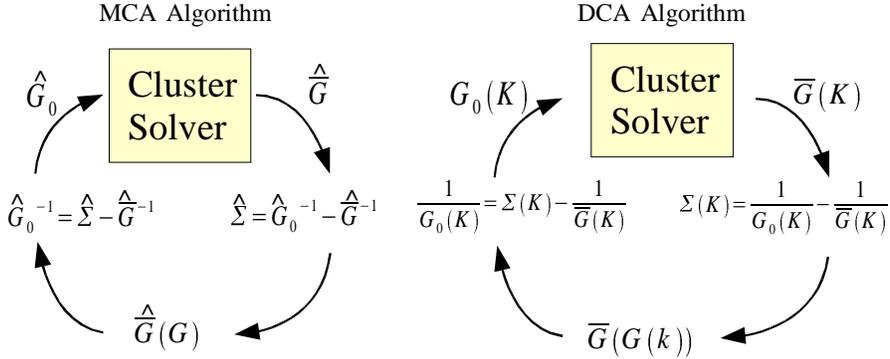}
\caption[]{The MCA and DCA algorithms.  The two differ mainly in that 
in the DCA, the Green functions are diagonal in $\K$, while in the
MCA, they are matrices in the cluster coordinates $\X$.}
\label{fig:algorithms}
\end{figure}

The MCA (DCA) algorithm, illustrated in Fig.~\ref{fig:algorithms},  follows 
directly: We first make an initial guess for the cluster self-energy matrix 
$\Sigma$.  This is used 
with Eqs.~\ref{eq:cgGMCA} and \ref{eq:GXk} (\ref{eq:cgGDCA} and \ref{G_DCA}) 
to calculate the coarse-grained Green function $\bar G$.  The 
cluster excluded Green function $\hat{\Gscr}=[\mbG^{-1}+\mS_{MC}]^{-1}$
(${\Gscr}(\K)= [\bG(\K)^{-1}+\Sigma_{DC}(\K)]^{-1}$)
is defined to avoid overcounting self energy corrections on the cluster.  
It is used to compute a new estimate for the cluster self-energy 
which is used to reinitialize the process.  

	Once convergence is reached, the irreducible quantities on the 
cluster may be used to calculate the corresponding lattice quantities.
For example, the cluster self energy and irreducible vertex functions
may be used in the Dyson and Bethe-Salpeter equations to calculate the
Green function and susceptibilities.  In order to obtain a smooth DCA 
self energy for the calculation of the band structure and spectra, it may 
be interpolated into the Brillouin zone of the lattice (however, such 
interpolation should be avoided during the self-consistency loop, as it 
may lead to causality violations).  The optimal MCA self energy, 
Eq.~\ref{eq:MDMFS} is a function of two momenta since the translational 
invariance of the lattice is violated.  Kotliar {\em{et al.}}, have 
introduced a cluster averaging scheme to obtain a self energy as a 
function of one momenta which may be employed after convergence is 
obtained \cite{kotliar}.

\subsection*{The Small Parameter, $\mGa$: the Coupling Between the Cluster 
and its Host}
In order to compare the character of the two different cluster approaches 
as a function of the cluster size $N_c$ it is instructive to rewrite the 
corresponding coarse grained Green-functions Eqs.~\ref{eq:cgGMCA} and 
\ref{eq:cgGDCA} to suitable forms by making use of the independence 
of the self-energy $\Sigma$ on the integration variable $\tk$. For the 
MCA coarse grained Green function we find
\begin{equation}
\label{eq:hMCA}
\mbG(z)=\left[zI-\hep_o-\mS_{MC}(z)-\mGa_{MC}(z)\right]^{-1}\,,
\end{equation}
with the ``cluster-local'' energy $\hep_o=N_c/N\sum_{\tk}\hat{\epsilon}(\tk)$. 
For the DCA we 
obtain a similar expression
\begin{equation}
\label{eq:hDCA}
\bG(\K,z)=\left[z-\bep-\Sigma_{DC}(\K,z)-\Gamma_{DC}(\K,z)\right]^{-1}\,,
\end{equation}
with the coarse grained average $\bep=N_c/N\sum_{\tk}\ep(\K,\tk)$.
The hybridization functions $\mGa_{MC/DC}(z)$ describe the coupling of the 
cluster to the mean-field representing the remainder of the system.

The behavior of $\Gamma$ for large $N_c$ is important. For the MCA, 
$\Gamma$ averaged over the cluster sites and frequency 
\begin{equation}
\label{eq:GbMCAm}
\bar{\Gamma}_{MC}=
\frac{1}{N_c}\sum_{\X_1,\X_2}\Gamma_{MC}(\X_1,\X_2)
\sim{\cal O}\left(\frac{2D}{L_c}\right)\,, 
\end{equation}
where $L_c=N_c^{1/D}$ is the linear cluster size.  A detailed derivation
of this form is presented in the appendix.  However, since in the MCA the 
cluster is defined in real space with open boundary conditions, this form 
is evident since only the sites on the surface $\propto 2D\cdot L_c^{D-1}$ 
of the cluster couple to the effective medium and $N_c=L_c^D$.  For the DCA we show in the appendix that $\Gamma(\K)\sim {\cal O}(1/N_c^{2/D})$ (see also \cite{DCA_Maier1}) 
so that we obtain for the average hybridization of the DCA cluster to the 
effective medium 
\begin{equation}
\label{eq:GavDCm}
\bar{\Gamma}_{DC}=
\frac{1}{N_c}\sum_{\K}\Gamma_{DC}(\K)\sim{\cal O}\left(\frac{1}{L_c^2}\right)\,.
\end{equation}
The DCA coarse graining results in a cluster in $\K$-space; thus, the 
corresponding real space cluster has periodic boundary conditions, and 
each site in the cluster has the same hybridization strength $\bar\Gamma$ 
with the host.

As shown in the appendix, the average hybridization strength $\bar \Gamma$ acts
as the small parameter in both the MCA and the DCA.    Thus the MCA (DCA) is an approximation 
with corrections of order $\bar{\Gamma}$ $\sim{\cal O}(1/L_c)$
($\sim{\cal O}(1/L_c^2)$).

\section*{Numerical Results}
To illustrate the differences in convergence with cluster size $N_c$
we performed MCA and DCA simulations for the symmetric one-dimensional 
(1D) Falicov-Kimball model (FKM). At half filling the FKM Hamiltonian reads
\begin{equation}
H=-t\sum_i (d_i^\dagger d_{i+1}^{}+h.c.)+U\sum_i(n_i^d-1/2)(n_i^f-1/2)\,,
\end{equation} 
with the number operators $n_i^d=d_i^\dagger d_i^{}$ and 
$n_i^f=f_i^\dagger f_i^{}$ and the Coulomb repulsion $U$ between $d$ 
and $f$ electrons residing on the same site.  The FKM can be considered as 
a simplified Hubbard model with only one spin-species ($d$) being allowed 
to hop. However it still shows a complex phase diagram including a Mott 
gap for large $U$ and half filling, an Ising-like charge ordering with the 
corresponding transition temperature $T_c$ being zero in 1D, and phase 
separation in all dimensions. The bare dispersion (in 1D) 
$\epsilon_k=2t\cos k$; thus for $t=1/4$ the bandwidth $W=1$ which we use 
as unit of energy. To simulate the effective cluster models of the MCA 
an the DCA we use a quantum Monte Carlo (QMC) approach described 
in \cite{DCA_Hettler2}. 

\begin{figure}[htb]
\includegraphics[width=.8\textwidth]{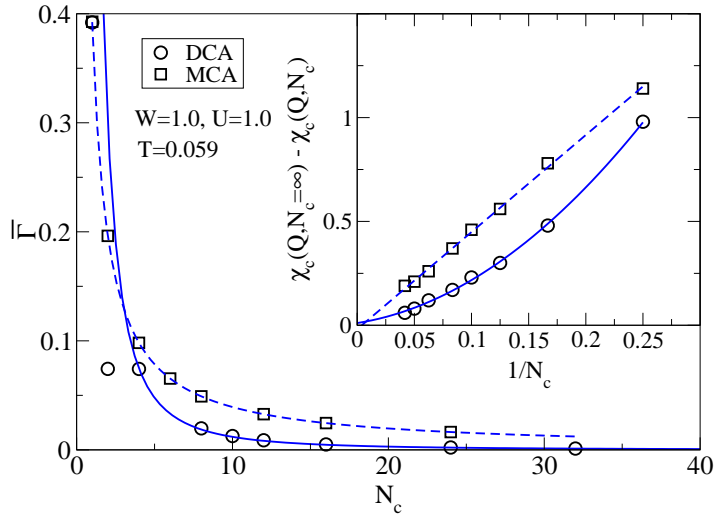}
\caption[]{The average integrated hybridization strengths $\bar{\Gamma}$ of 
the MCA (squares) and DCA (circles) versus the cluster size $N_c$ when 
$\beta=17$ and $U=W=1$ in the symmetric model. The solid and dashed lines represent the fits 
$1.1946/N_c^2$ and $0.3361/N_c$ respectively. Inset: Convergence of the 
cluster charge susceptibility for $Q=\pi$. The solid and dashed lines 
are quadratic and linear fits, respectively.}
\label{fig:GvsL}
\end{figure}

To check the scaling relations Eqs.~\ref{eq:GbMCAm} and \ref{eq:GavDCm}, 
we show in Fig.\ref{fig:GvsL} the average hybridization functions 
$\bar{\Gamma}_{MC}$ and $\bar{\Gamma}_{DC}$ for the MCA and DCA respectively 
at the inverse temperature $\beta=17$ for $U=W=1$.   
For $N_c=1$ both approaches are equivalent to the DMFA and thus 
$\bar{\Gamma}_{MC}=\bar{\Gamma}_{DC}$. For increasing $N_c$ 
$\bar{\Gamma}_{MC}$ can be fitted by $0.3361/N_c$ and $\bar{\Gamma}_{DC}$ 
by $1.1946/N_c^2$ when $N_c>2$. 
Cluster quantities, such as the self energy and cluster susceptibilities, 
are expected to converge with increasing $N_c$ like $\bar{\Gamma}$.  This 
is illustrated in the inset for the staggered ($Q=\pi$) charge 
susceptibility $\chi_c({\bf Q})$ of the cluster.

Since only the compact parts represented by $\Phi$ of the lattice free energy 
(Eq.~\ref{F_CA}) are coarse-grained, this scaling is expected to break down 
when lattice quantities, such as the lattice charge susceptibility, are 
calculated.  The susceptibility of the cluster $\chi_c(Q)$ cannot 
diverge for any finite $N_c$; whereas the lattice $\chi(Q)$ diverges 
at the transition temperature $T_c$ to the charge ordered phase.  Note that 
the residual mean-field character of both methods can result in finite 
transition temperatures $T_c>0$ for finite $N_c<\infty$. However as $N_c$ 
increases, this residual mean field character decreases gradually and thus 
increased fluctuations should drive the solution to the exact result $T_c=0$.

\begin{figure}[htb]
\includegraphics[width=.8\textwidth]{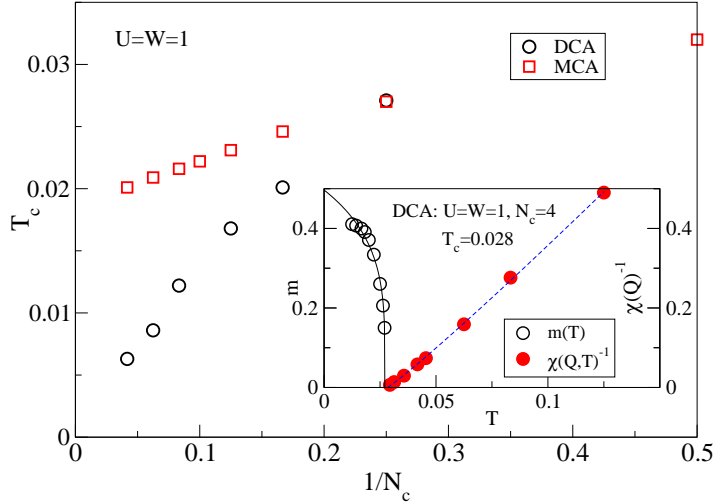}
\caption[]{The transition temperature $T_c$ for the DCA (circles) and 
MCA (squares) when $U=W=1$ versus the cluster size $N_c$. For all 
values of $N_c$ the DCA prediction is closer to the exact result 
($T_c=0$). Inset: Order parameter $m(T)$ and inverse charge 
susceptibility $\chi(Q)^{-1}$ versus temperature. The solid (dashed) line 
represents a  fit to the functions $m(T)\propto (T_c-T)^\beta$ with 
$\beta=0.245$ ($\chi(T)\propto (T-T_c)^{-\gamma}$ with $\gamma=1.07$).}
\label{fig:phase}
\end{figure}

In the DCA \cite{DCA_Hettler2}, $\chi(Q)$ is calculated by first extracting 
the corresponding vertex function from the cluster simulation.  This is then 
used in a Bethe-Salpeter equation to calculate $\chi(Q)$.  $T_c$ is 
calculated by extrapolating $\chi(Q)^{-1}$ to zero using the function 
$\chi(Q)^{-1}\propto (T-T_c)^\gamma$ (see inset to Fig.\ref{fig:phase}).
This procedure is difficult, if not impossible, in the MCA due to the lack 
of translational invariance. Here, we calculate the order parameter 
$m(T)=1/N_c\sum_i(-1)^i\langle n_i^d\rangle$ in the symmetry broken phase.  
$T_c$ is then obtained from extrapolating $m(T)$ to zero using the function 
$m(T)\propto (T_c-T)^\beta$. For the DCA this extrapolation is shown by the 
solid line in the inset to Fig.\ref{fig:phase} for $N_c=4$.  The values for 
$T_c$ obtained from the calculation in the symmetry broken phase and in the 
unbroken phase must agree, since as we have shown above, both the DCA 
and MCA are $\Phi$-derivable.  This is illustrated in Fig.\ref{fig:phase} for 
the DCA.    

A comparison of the DCA and MCA estimate of $T_c$ is presented in
Fig.~\ref{fig:phase}.  $T_c$ obtained from MCA (squares) is larger than 
$T_c$ obtained from DCA (circles). Moreover we find that the DCA result 
seems to scale to zero almost linearly in $1/N_c$ (for large enough $N_c$), 
whereas the MCA does not show any scaling form and in fact seems to tend to 
a finite value for $T_c$ as $N_c\rightarrow\infty$. This striking 
difference of the two methods can be attributed to the different 
boundary conditions. The open boundary conditions of the MCA cluster result 
in a large surface contribution so that $\bar{\Gamma}_{MC}>\bar{\Gamma}_{DC}$.  
This engenders pronounced mean field behavior that stabilizes the finite 
temperature transition for the cluster sizes treated here. For larger 
clusters we expect the bulk contribution to the MCA free energy to dominate 
so that $T_c$ should fall to zero. 

Complementary results are found in simulations of {\em{finite-sized}} 
systems.  In general, systems with open boundary conditions are expected 
to have a surface contribution in the free energy of order 
${\cal{O}}(1/L_c)$ \cite{M_Fisher_72}.   This term is absent in systems 
with periodic boundary conditions. As a result, simulations of 
finite-sized systems with periodic boundary conditions converge much 
more quickly than those with open boundary conditions \cite{D_Landau_76}. 

\begin{figure}[htb]
\includegraphics[width=1.0\textwidth]{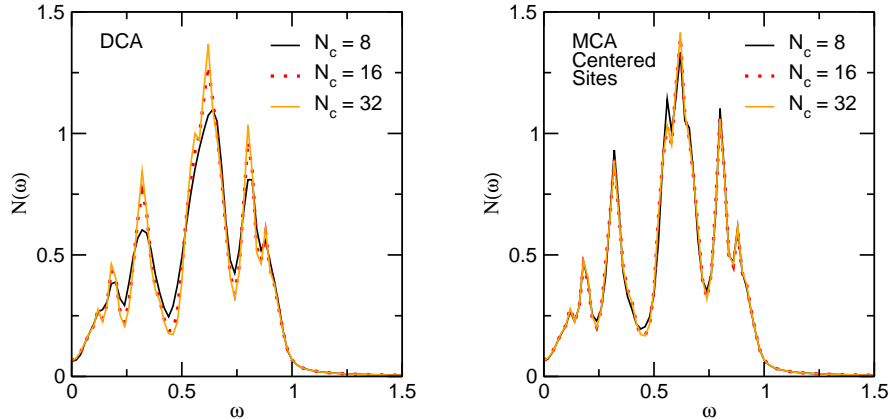}
\caption[]{Comparison of the single-particle density of states (DOS)
calculated with the DCA (left) and the MCA (right) on the two
central sites for various cluster sizes $N_c$ when $\beta=10$ and 
$U=W=1$ (since the DOS is symmetric around $\omega=0$, only the 
$\omega\geq 0$ part is shown).  Local quantities, such as the DOS, 
converge more quickly when calculated on the central sites of the 
MCA cluster than the cluster averaged DCA result.}
\label{fig:dos_DCA_MCAcenter}
\end{figure}

Thus far, we have shown that the DCA converges more quickly than the 
MCA for critical properties and for extended cluster quantities 
(e.g.\ the cluster susceptibility).  This is due to
the differences in the boundary conditions, and the coupling to
the mean-field.  Whereas each site in the DCA experiences the same 
coupling to the mean-field host, in the MCA only the sites on the 
boundary of the cluster couple to the host.  Provided that the 
system is far from a transition, the sites in the center of the cluster 
couple to the mean-field only through propagators which fall 
exponentially with distance.  Thus, one might expect that local results, 
such as the single-particle density of states, might converge more 
quickly within the MCA provided that they are measured on these central 
sites.  This is illustrated in Fig.~\ref{fig:dos_DCA_MCAcenter} where we 
plot the single-particle density of states calculated with the DCA and 
the MCA on the two central sites.

\section*{Summary.}  By defining appropriate Laue functions, we
provide microscopic diagrammatic derivations of the MCA and DCA.   We 
show that they are $\Phi$-derivable, and that the lattice free energy
is optimized by equating the irreducible quantities on the lattice to 
those on the cluster.  The MCA maps the lattice to a cluster with open 
boundaries and consequently, the cluster violates translational 
invariance.  In contrast, the DCA cluster has periodic boundary 
conditions, and therefore preserves the translational invariance of 
the lattice. 

	These differences in the boundary conditions translate
directly to different asymptotic behaviors for large clusters $N_c$. 
As we find analytically as well as numerically, the surface 
contributions in the MCA lead to an average hybridization 
$\bar\Gamma$ of the cluster  to the mean field that scales like 
$1/L_c$ as compared to the $1/L_c^2$ scaling of the DCA.   Since 
$\bar\Gamma$ acts as the small parameter for these approximation 
schemes, the DCA converges much more quickly than the MCA.  These 
effects are more pronounced near a transition, where the large 
surface contribution of the MCA stabilizes the mean-field character
of the transition.  Consequently, the DCA result for the transition 
temperature $T_c$ of the 1D symmetric FKM model scales almost like 
$1/N_c$ to the exact result $T_c=0$, whereas the MCA result converges 
very slowly. The boundary conditions also differ in that only the MCA
sites at the surface of the cluster couple to the mean field; 
whereas, all DCA cluster sites have an equal coupling to the mean
field host.  As a result local quantities, such as the density of
states, when measured on the central sites of the MCA cluster converge 
more quickly than than corresponding DCA results.  

Thus, for critical properties and extended cluster quantities, the
DCA converges far more quickly than the MCA; whereas for local quantities
which may be measured at the central sites of the MCA cluster, the
MCA converges more quickly.  Since the origin of these differences 
lies in the different boundary conditions we expect them to hold 
generally for any model of electrons moving on a lattice.

\paragraph*{Acknowledgements} We acknowledge useful conversations with 
N.\ Bl\"umer,
A.\ Gonis,
M.\ Hettler,
H.R.\ Krishnamurthy,
D.P.\ Landau,
Th.\ Pruschke,
W.\ Shelton,
and 
A.\ Voigt. 
This work was supported by NSF grants DMR-0073308 and 0113574.  This 
research was supported in part by NSF cooperative agreement ACI-9619020 
through computing resources provided by the National Partnership for 
Advanced Computational Infrastructure at the Pittsburgh Supercomputer 
Center. 

\appendix

\section*{Appendix}

To differentiate the MCA from the DCA we find expressions for the 
corresponding host-functions $\Gamma$. To this end we split the hopping 
integral $t$ into an inter- and intracluster part. For the MCA the 
intercluster hopping $[\mT_{MC}]_{\X_1,\X_2}=T_{MC}(\X_1,\X_2)$ is defined 
as $\mT_{MC}(\tk)=\mep(\tk)-\hep_o$ and reads in real space 
\begin{equation}
\mT_{MC}(\tx)=\mt(\tx)-\hep_o\de (\tx)\,.
\end{equation}
Since $\hep_o=\mt(\tx=0)$, the intercluster 
hopping matrix $\mT_{MC}(\tx)$ is only finite for $\tx\neq 0$ as expected. 
Thus, $\mT_{MC}(\tx)$ has non-vanishing matrix-elements only for sites on 
the boundary of the cluster. For the DCA the intercluster hopping 
integral is analogously defined as $T_{DC}(\K,\tk)=\ep(\K,\tk)-\bep$ and 
can be written in real space as
\begin{equation}
\mT_{DC}(\tx)=\mt(\tx)-\hat{\bar{t}}\hat{\de}^c(\tx)\,,
\end{equation}  
where $[\hat{\bar{t}}]_{\X_1\X_2}=\bar{t}(\X_1-\X_2)=
1/N_c\sum_{\K}e^{i\K\cdot(\X_1-\X_2)}\bep$ and the cluster delta-like function $[\hat{\de}^c(\tx)]_{\X_1\X_2}=\de^c(\X_1-\X_2,\tx)=
N_c/N\sum_{\tk}e^{i\tk\cdot(\X_1-\X_2+\tx)}$. We have discussed elsewhere 
that $\de^c(\X_1-\X_2,\tx)\approx 1$ for $\tx=0$ and falls off rapidly 
for finite $\tx$. It is important to note that $\bar{t}(\X_1-\X_2)=
t(\X_1-\X_2,\tx=0)$ for $N_c=N$, but $|\bar{t}(\X_1-\X_2)|<|t(\X_1-\X_2,\tx=0)|$ 
for $N_c<N$.  It follows that the DCA intercluster hopping $T_{DC}(\X_1-\X_2,\tx)$ 
for $\tx=0$, i.e. between sites belonging to the same cluster, is even finite 
when $N_c<N$; however it is strongly reduced compared to $\tx\neq 0$. This is 
a consequence of the periodicity of the DCA cluster. Therefore we find that 
the effective model of the DCA cannot be thought of $N/N_c$ clusters composing 
the original lattice, but we have to rather think of it as a renormalized model 
in real space.

Now we proceed with finding expressions for the hybridization functions 
$\Gamma$ defined in Eq.~(\ref{eq:hMCA}) and (\ref{eq:hDCA}) for the 
MCA and DCA. We start with the MCA. To this end we define a Green function 
matrix
\begin{equation}
\mg(z)=\left[zI-\hep-\mS_{MC}(z)\right]^{-1}\,,
\end{equation}
that is localized on the impurity cluster. With this definition 
Eq.~(\ref{eq:cgGMCA}) reads
\begin{equation}
\label{eq:cGalt}
\mbG=\frac{N_c^2}{N^2}\sum_{\tk}\mG(\tk)=\left[\mg^{-1}-\mGa_{MC}\right]^{-1}\,,
\end{equation}
and Eq.~(\ref{eq:GXk}) can be rewritten to
\begin{equation}
\label{eq:Galt1}
\mG(\tk)=\mg+\mg\mT_{MC}(\tk)\mG(\tk)\,.
\end{equation}
After inserting Eq.~(\ref{eq:Galt1}) into Eq.~(\ref{eq:cGalt}) and using $\sum_{\tk}\mT_{MC}(\tk)=0$ we obtain after some algebraic transformations for the MCA-hybridization matrix
\begin{eqnarray}
\mGa_{MC}&=&\left[I+\frac{N_c}{N}\sum_{\tk}\mT_{MC}(\tk)\mG(\tk)\right]^{-1}\times\nonumber\\
&&\frac{N_c}{N}\sum_{\tk}\mT_{MC}(\tk)\mG(\tk)\mT_{MC}(\tk)\,.
\end{eqnarray} 
Since, as we pointed out above, $\mT$ has non-zero matrix-elements only 
between sites on the boundary of the impurity cluster, $\bar\Gamma$ is finite 
only for sites on the boundary but vanishes for sites inside the cluster. 
Thus, the average hybridization strength of the MCA cluster per site 
\begin{equation}
\label{eq:GbMCA}
\bar{\Gamma}_{MC}=\frac{1}{N_c}\sum_{\X_1,\X_2}\Gamma_{MC}(\X_1,\X_2)\sim{\cal O}\left(\frac{2D}{L_c}\right)\,, 
\end{equation}
since only the sites on the surface $\propto 2D\cdot L_c^{D-1}$ of the 
cluster couple to the effective medium and $N_c=L_c^D$.

For the DCA we can follow the steps presented above for the MCA since 
$\Gamma_{DC}(\K)$ can be considered a diagonal matrix in $\K$. We obtain for 
the DCA hybridization function
\begin{equation}
\Gamma_{DC}(\K)=\frac{\displaystyle \frac{N_c}{N}\sum_{\tk}T_{DC}^2(\K,\tk)G(\K,\tk)}{\displaystyle 1+\frac{N_c}{N}\sum_{\tk}T_{DC}(\K,\tk)G(\K,\tk)}\,.
\end{equation}  
It should be stressed that due to the periodicity of the DCA cluster every 
site of the impurity cluster couples to the effective medium. The momenta 
$\tk$ are restricted to a DCA coarse graining cell and therefore maximal 
of the order ${\cal O}(\Delta k)$, where $\Delta k\propto 1/L$. Since 
$\ep({\bf{K}})-\bar{\epsilon}_{\bf K}\sim{\cal O}((\Delta k)^2)$, we find by performing a Taylor series 
expansion of $T_{DC}(\K,\tk)$ around $\ep(\K)$ that 
$T_{DC}(\K,\tk)\sim{\cal O}(1/L_c)$.  For the overall hybridization of the DCA 
cluster to the effective medium we thus obtain 
\begin{equation}
\bar{\Gamma}_{DC}=\frac{1}{N_c}\sum_{\K}\Gamma_{DC}(\K)\sim{\cal O}\left(\frac{1}{L_c^2}\right)\,.
\end{equation}

In both the DCA and the MCA, the average hybridization strength acts
as the small parameter.  The approximation performed by the DCA (MCA) is 
to replace the lattice Green function $G(\K,\tk,z)=[z-\ep_{\K+\tk}-\Sigma(\K,\tk,z)]^{-1}$ ($\mG(\tk)=[zI-\hep(\tk)-\mS(\tk,z)]^{-1}$) by its 
coarse grained quantity $\bG(\K)$ ($\mbG$) 
in diagrams for the generating functional $\Phi$. 

According to Eq.~(\ref{eq:hDCA}) the coarse grained Green function of the 
DCA can be expressed as 
$\bG(\K,z)=\left[z-\bep-\Sigma_{DC}(\K,z)-\Gamma_{DC}(\K,z)\right]^{-1}$.  
For the time being, assume that $\Sigma_{DC}(\K,z)$ has corrections of 
the same, or higher, order in $1/L_c$ as the average hybridization strength.  
Both the self energy and $\epsilon_{\K+\tk}=\bep+{\cal O}(1/L_c)$ with the 
leading order corrections being linear in $\tk$. Since furthermore 
$\Gamma_{DC}(\K)\sim{\cal O}(1/L_c^2)$, 
$G(\K,\tk)\approx \bG(\K)+{\cal O}(1/L_c)$. The diagrams for $\Phi$ however 
are summed over $\tk$, so that the terms $\sim {\cal O}(1/L_c)$ coming from 
$\epsilon_{\K+\tk}$ and similar terms from the self energy vanish and only 
the terms $\sim {\cal O}(1/L_c^2)$, or higher, survive. Thus we find for the 
DCA that $\Phi\approx\Phi_{DC}+{\cal O}(1/L_c^2)$.  The corresponding estimate 
of $\Sigma_{DC}(\K,z)$ will also have corrections ${\cal O}(1/L_c^2)$, 
confirming the assumption above.

According to Eq.~(\ref{eq:hMCA}), the coarse grained propagator of the 
MCA can be written as 
$\mbG(z)=[zI-\hat{\ep}_o-\mS_{MC}(z)-\mGa_{MC}(z)]^{-1}$. 
If we assume that $\mS_{MC}$ converges with $L_c$ as fast or faster than $1/L_c$  
we see with the $L_c$-dependence of $\mGa_{MC}$ in Eq.~(\ref{eq:GbMCA}) 
that $\mG(\tk)\approx\mbG+{\cal O}(1/L_c)$ since 
$\hat{\ep}(\tk)=\hep_o+{\cal O}(1/L_c)$. Thus we  obtain  for the MCA 
approximation $\Phi\approx\Phi_{MC}+{\cal O}(1/L_c)$ and $\mS$ will 
converge with $L_c$ as ${\cal O}(1/L_c)$ confirming our assumption.

%

\end{document}